\documentclass[twocolumn, pra, aps, superscriptaddress,longbibliography,showpacs,amsmath,amssymb,floatfix]{revtex4-1}                   
\usepackage{changes}
\usepackage[normalem]{ulem}  

\usepackage{graphicx}
\usepackage{dcolumn}
\usepackage{bm}
\usepackage{graphicx}                                               
\usepackage{amssymb}

\usepackage{amsmath}
\usepackage{epsfig}
\usepackage{xcolor}
\usepackage{tabu}
\usepackage{mathtools}
\usepackage[colorlinks,linkcolor=blue,anchorcolor=blue,citecolor=blue,urlcolor=blue]{hyperref}
\usepackage{physics}
\usepackage{float}
\usepackage{diagbox}
\usepackage{inputenc}

\begin{document}
\title{Soliton and traveling wave solutions in 
 coupled one-dimensional condensates}
\author{Zeyu Rao}
\affiliation{Key Laboratory of Quantum Information, University of Science and Technology of China, Hefei 230026, China}
\author{Xiaoshui Lin}
\affiliation{Key Laboratory of Quantum Information, University of Science and Technology of China, Hefei 230026, China}
\author{Jingsong He}
\affiliation{Institute for Advanced Study, Shenzhen University, Shenzhen, 518000, China}
\affiliation{College of Physics and Optoelectronic Engineering, Shenzhen University, 518060, Shenzhen, China}
\author{Guangcan Guo}
\affiliation{Key Laboratory of Quantum Information, University of Science and Technology of China, Hefei 230026, China}
\affiliation{Hefei National Laboratory, University of Science and Technology of China, Hefei 230088, China}
\affiliation{Synergetic Innovation Center of Quantum Information and Quantum Physics, University of Science and Technology of China, Hefei 230026, China}
\affiliation{Anhui Province Key Laboratory of Quantum Network,
University of Science and Technology of China, Hefei 230026, China}
\author{Ming Gong}
\email{gongm@ustc.edu.cn}
\affiliation{Key Laboratory of Quantum Information, University of Science and Technology of China, Hefei 230026, China}
\affiliation{Hefei National Laboratory, University of Science and Technology of China, Hefei 230088, China}
\affiliation{Synergetic Innovation Center of Quantum Information and Quantum Physics, University of Science and Technology of China, Hefei 230026, China}
\affiliation{Anhui Province Key Laboratory of Quantum Network,
University of Science and Technology of China, Hefei 230026, China}

\date{\today }

\begin{abstract}
Ultracold condensates provide a unique platform for exploring soliton physics. Motivated by the recent experiments realizing the sine-Gordon model in a split one-dimensional (1D) BEC, we demonstrate that this system naturally supports various density and phase solitons. We explore the physics using the bosonization technique, in which the phase and density are conjugate pairs, and determine its effective Language equation and the associated equation of motion. We show that in the presence of asymmetry between the two condensates, new solutions beyond those in the sine-Gordon model emerge. We calculate the traveling wave solutions and soliton solutions in this model and determine their corresponding energy densities analytically. Finally, we discuss the relevance of these solutions to the experiments and discuss their observations. This theory does not rely on the mechanism of quasi-particle excitation, which yields the Lee-Huang-Yang correction in higher dimensions, and is thus much more suitable to describe the physics in  1D systems. Since the physical models have already been realized in experiments, this work opens a new frontier for the realization of various soliton and periodic solutions using two coupled condensates. 
\end{abstract}

\maketitle 

The dynamics of one-dimensional (1D) Bose-Einstein condensate (BEC) can be described by the following Gross-Pitaevskii (GP) equation 
 $i\hbar{\partial \psi (x, t)\over \partial t} = (-{\frac{\hbar^{2}}{2m}{\partial^2 \over \partial x^2} + g|\psi(x, t)|^2}) \psi(x, t)$ \cite{Pitaevskii2016BEC,Tsuzuki1971Nonlinear,Morsch2006Dynamics,Dalfovo1999TheoryBEC,Konotop2004Landau} , where $m$ is the atom mass, $\hbar$ is the reduced Planck constant and $g$ is the many-body interaction strength. To some extent, this equation can be regarded as exact from the many-body interacting model  \cite{Laszio2007Rigous, Lieb2000BosonsRigorous, Lieb2022Proof, Dalfovo1999TheoryBEC, Gross1963Hydrodynamics}. It has been well-known that this model may support two kinds of solitons, that is, the dark soliton and bright state \cite{Salasnich2004Dynamics, drazin1989solitons, shabat1972exact, Jason2017Formation,  Christoph2008Oscillations}. Let $\psi(x,t)=\phi(x)e^{-i {\mu \over \hbar}t}$, then we have $(-{\hbar^2 \over 2m} {\partial^2 \over \partial x^2} + g |\phi(x)|^2) \phi(x) = \mu \phi(x)$. When $g < 0$, it admits a self-focusing bright soliton as
\begin{equation}
\phi(x) = {\sqrt{m|g|}\over 2\hbar} \sech({m|g| x \over 2\hbar^2}), \quad \mu = -{mg^2 \over 8 \hbar^2}.
\label{eq-brightsoliton}
\end{equation}
In this case, the uncertainty principle prohibits the collapse of the condensate, leading to a soliton with finite width. The details of this solution are presented in supplemental material  S-I \cite{supplematRao}. For $g > 0$ it admits a self-defocusing dark soliton as 
\begin{equation}
\phi(x) = A \cdot \tanh(\sqrt{{mg \over \hbar^2}} Ax), \quad \mu = g A^2,
\label{eq-darksoliton}
\end{equation}
where $A$ is an arbitrary constant. These solitons have been realized in recent experiments \cite{Burger1999DarkSolitons,Nguyen2014Collisions,Khaykovich2002Formation,Medley2014Evaporative,Becker2008Oscillations,Eiermann2004Bright,Bennet2011Observation,Denschlag2000Generating,strecker2002formation,Anderson2001Watching,Weller2008Experimental, Sanz2022Interaction, Sean2024Observation}. In the presence of velocity, these solutions need to be changed, in which $\psi$ may contain both real and imaginary parts, yielding much more complicated solitons. Beyond these exact solvable conditions, the analytical forms of these solutions can not be obtained,  which could be handled using the variational principle \cite{PerezGarcia1997Dynamics, PerezGarcia1996LowEnergy,Salasnich2002Effective}. The ansatz in this approach has limited our understanding of the soliton physics of 1D condensates.

Strictly speaking, these solutions can not fully reflect the unique feature of 1D systems, in which any particles, including bosons, fermions, and anyons, could be described by collective excitations in terms of bosonization and renormalization group (RG) method \cite{Haldane1981effective, giamarchi2003quantum, Shankar1994Renormalization}. Thus, these different particles should have the same physics. In this theory, the phase field and density field are canonical conjugate pairs; thus, any spatial structures in the phases should be manifested from their densities. This requirement should also be reflected in the theoretical investigation of the excitation in these models, in which one may assume the possible soliton solution and solve several coupled nonlinear equations based on the Lagrange equation \cite{Busch2001DarkBright,Liang2005Dynamics,Li2005Exact}. Their general analytic forms can not be obtained in most cases, which can be studied using the numerical simulations method \cite{Pedri2005TwoDimensional,Xu2023GapSoliton,Lobanov2014Fundamental,Yang2023SOC, Zhang2009Gap,Ahufinger2005Lattice,Kartashov2013Gap}. Up to date, only a few of these models with tailored interaction in the 1D BEC can be solved analytically \cite{Hirota1971Exact,LiLu2006Exact, Yu2022Propagating,ZhaoLiChen2020Magnetic,LiLu2006Exact,Ieda2004Exact,Mao2022Exact,QuChunlei2016Magnetic}.

In this manuscript, we use the bosonization technique \cite{Haldane1981effective,giamarchi2003quantum} to study the nonlinear dynamics and related soliton physics in two coupled 1D condensates. Firstly, we show that in the true 1D system with density fluctuation, in which the phase and density are conjugate pairs, the physics could be described by collective excitations. Next, we show that the effective Lagrangian is given by the generalized 
sine-Gordon (sG) model \cite{Khusnutdinova2003On,RAY2006A,Sadighi2009TRAVELINGWS,SALAS2010Exact, zhao2011exact,qin2017the, EKICI2017Exact,ILATI2015the}. We study various solutions in this model, including the phase solitons and the conjugate density solitons. Finally, we estimate their observation using experimental parameters. Since the two-component model has been intensively studied with ultracold atoms in experiments \cite{haller2010pinning,schweigler2017experimental,schweigler2021decay, Farolfi2020Observation}, and the soliton physics in coupled sine-Gordon equations has not yet been fully understood, the effective model derived in this work should provide a fertile ground for new nonlinear physics in future experiments, including various traveling wave and soliton solutions. 

\begin{figure}[htbp]
\centering
\includegraphics[width=0.48\textwidth]{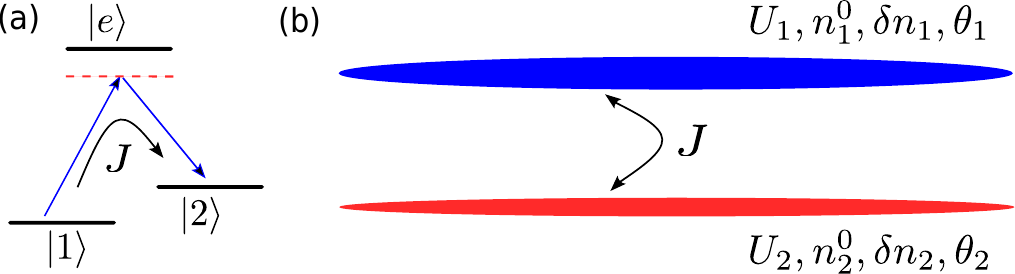}
\caption{(a) Coupling of two condensates from stimulated Raman transition in a three-level system. (b) Two coupled one-dimensional condensates with asymmetry parameters; see their physical meanings in the main text. }
\label{fig-fig1}
\end{figure}
 
We start from the Lagrange in the coupled condensate realized in recent experiments \cite{haller2010pinning,schweigler2017experimental,schweigler2021decay} (see Fig. \ref{fig-fig1})
\begin{equation}
\mathcal{L} = \sum_{i=1,2} {i\hbar \over 2} (\psi_i^\dagger \partial_t \psi_i - \psi_i \partial_t \psi_i^\dagger) - \mathcal{H},
\label{eq-effectiveL}
\end{equation}
with $\mathcal{H} = \sum_{i=1,2} \psi_i^\dagger ({p^2 \over 2m} -\mu_i)\psi_i + {U_i \over 2} (\psi_i^\dagger \psi_i)^2 + J\psi_1^\dagger \psi_2 + \text{h.c.}$,
where $\psi_i$ is the bosonic field for the $i$-th chain/condensate with $m$ being the particle mass, $\mu_i$ is the chemical potential, $U_i$ is the many-body interaction,
and $J$ is the hopping strength between the two condensates. $\mathcal{H}$ and $\mathcal{L}$ are the densities of the Hamiltonian and Lagrangian, respectively. The two states $|\psi_1\rangle$ and $|\psi_2\rangle$  can be realized using the hyperfine states in the ultracold atoms, and their coupling $J$ can be realized by the Raman coupling; see Fig. \ref{fig-fig1}(a). In terms of density-phase representation $\psi_i = e^{i\theta_i}\sqrt{n_i^0 + \delta n_i}  $ \cite{Cazalilla2011one,Haldane1981effective,Ruggiero2021LargeScale, giamarchi2003quantum, rao2023generalized}, 
where $n_i^0$ is the mean density and $\delta n_i$ is its fluctuation,
and $\theta_i$ is the phase field with quantization condition $[\delta n_i(x,t), \theta_j(x',t)] = i\delta_{ij}\delta(x-x')$. After integrate out the density field $\delta n_i$, we can obtain an effective Lagrangian \cite{rao2023generalized}
\begin{equation}
	\mathcal{L} = {\hbar^{2}\over 2} \sum_i  ({(\partial_t \theta_i)^2 \over U_i} - \frac{n_{i}^{0}}{m} (\partial_x \theta_i)^2) - \hbar g \cos(\theta_1 - \theta_2),
\end{equation}
where $K_i = \hbar\sqrt{{n_i^0 \over m U_i}}$, $u_i = \sqrt{{n_i^0 U_i  \over m}}$, and  $g = {2J\over \hbar}  \sqrt{n_1^0 n_2^0}$.  These three parameters can be controlled in experiments by controlling the population imbalance, Raman coupling, and interaction strengths, therefore, we will treat them as independent variables. 

The corresponding equation of motion can be obtained \cite{Landau1976Mechanics} 
\begin{eqnarray}
    && \frac{K_{1}}{u_{1}}\partial_{t}^{2}\theta_{1}-K_{1}u_{1}\partial_{x}^{2}\theta_{1}=g\sin(\theta_{1}-\theta_{2}), \\
    && \frac{K_{2}}{u_{2}}\partial_{t}^{2}\theta_{2}-K_{2}u_{2}\partial_{x}^{2}\theta_{2}=-g\sin(\theta_{1}-\theta_{2}).
\end{eqnarray}
In terms of the density-phase conjugate pairs, we have 
\begin{equation}
    \delta n_{i}=-\frac{\hbar\partial_{t}\theta_{i}}{U_{i}}.
    \label{eq-conjugatepairs}
\end{equation}
See detailed derivation in supplemental material S-II \cite{supplematRao}. This relation means that if $\theta_i$ exhibits some spatial structure (including soliton solutions), so does $\delta n_i$, and vice versa. The above conjugate relation is one of the major starting points of this manuscript. From this constraint, we see that any spatial structure in the phase can induce corresponding structures in its density. It should be emphasized that this is a standard relation in bosonization \cite{Gritsev2007Linear, Whitlock2003Relative}. 
Further, with the transformation $t^{\prime}=\sqrt{\frac{ gu_{1}}{K_{1}}}t$, and $x^{\prime}=\frac{1}{u_{2}}\sqrt{\frac{ gu_{1}}{K_{1}}}x$, we have the following coupled sine-Gordon equations 
\begin{eqnarray}
    \partial_{t^{\prime}}^{2}\theta_{1}- \alpha^2 \partial_{x^{\prime}}^{2}\theta_{1} &=&\sin{(\theta_{2}-\theta_{1})}, \label{eq-motion1} \\  
    \partial_{t^{\prime}}^{2}\theta_{2}-\partial_{x^{\prime}}^{2}\theta_{2} & =& -\delta^2 \sin{(\theta_{2}-\theta_{1})},
    \label{eq-motion2}
\end{eqnarray}
which contains only two induced parameters: $\alpha=u_{1}/u_{2}$, and $\delta^{2}=\frac{K_{1}u_{2}}{K_{2}u_{1}}$. Hereafter, we will drop the primes and let $t\rightarrow t^{\prime}$, $x\rightarrow x^{\prime}$ during the 
analytic calculation of the solutions of $\theta_i$ and $\delta n_i$, unless specified explicitly. 

The major task of this manuscript is to calculate the possible solutions of the above two phase equations. When these two chains have identical velocities and Luttinger parameters, the relative phase between the two chains will give the standard sine-Gordon equation, which has been realized in experiments \cite{haller2010pinning, schweigler2017experimental, schweigler2021decay}. It will lead to new solutions (hence new physics) when they are different (see Fig. \ref{fig-fig1}(b)). For this reason, in this work, the above two equations will be termed as the generalized sine-Gordon model, following Ref. \cite{rao2023generalized}. The total energy density of this model, turned back to the Hamiltonian form, can be split into two terms 
\begin{eqnarray}
    \mathcal{T} && = \sum_{i=1}^2  \frac{(\partial_{x}\delta n_{i})^{2}}{8 mn_{i}^{0}}+\frac{n_{i}^{0} (\partial_{x}\theta_{i})^{2}}{2 m}+\frac{1}{2}U_{i}\delta n_{i}^{2}, \label{eq-T} \\ 
    \mathcal{V} && =2J\sqrt{n_{1}^{0}n_{2}^{0}}\cos{(\theta_{1}-\theta_{2})} \label{eq-V}. 
\end{eqnarray}
as discussed in detail in supplemental material S-II \cite{supplematRao}. The Gibbs term proportional to $\delta n_i$ becomes a constant after integration and is therefore ignored. The total Hamiltonian density of this model is given by $\mathcal{\epsilon} = \mathcal{T} + \mathcal{V}$. 
Similarly, the total energy of each term can be written as $E= T+V$, 
with $T= \int_{-\infty}^{\infty}\mathcal{T}dx$, 
$V= \int_{-\infty}^{\infty}\mathcal{V}dx$. 

We notice that the two equations are standard coupled sine-Gordon equations introduced by Khusnutdinova and Pelinovsky \cite{Khusnutdinova2003On}, which can be regarded as a generalized form of the Frenkel–Kontorova dislocation model \cite{BRAUN1998Nonlinear}. It has also found applications in the DNA double-chain model \cite{Yomosa1983soliton}. Thus, the coupled sine-Gordon equations have been studied in different contexts 
\cite{RAY2006A,Sadighi2009TRAVELINGWS,SALAS2010Exact, zhao2011exact,qin2017the, EKICI2017Exact,ILATI2015the}, giving rise to various physics. This work mainly focuses on their possible soliton and traveling wave solutions in this new model. We focus on the two most transparent solutions in these equations and discuss their density and phase densities in terms of the conjugate relation in Eq. \ref{eq-conjugatepairs} and determine their energy density profiles. It should be noted that we have not exhausted all the solutions in this generalized sine-Gordon model, which are hoped to be explored using some more general mathematical tools. 

(I) {\it Coherent periodic traveling wave solution}. Firstly, we will show how this generalized sine-Gordon model is reduced to solutions in the sine-Gordon model with traveling wave solutions \cite{Fu2005exact,LIU2006exact,CHEN2008exact}. The solutions to the sine-Gordon equation are discussed in supplemental material S-III  \cite{supplematRao}. To this end, let us assume ($\xi = x -ct$)
\begin{equation}
\theta_1(x, t) = (q+1) \theta(\xi), \quad  \theta_2(x, t) = q \theta(\xi),
\end{equation}
where $c$ is the velocity of the traveling wave and $q$ is the ratio between these two fields. We find 
\begin{equation}
q = {K_1 u_2 (c^2 - u_1^2) \over u_1 u_2 (K_1 u_1 + K_2 u_2) -(K_2 u_1 + K_1 u_2)c^2},
\label{eq-q}
\end{equation}
and 
\begin{equation}
\theta''(\xi) = {1 \over A} \sin(\theta), 
\label{eq-thetapp}
\end{equation}
with $A =  (q+1) ((K_1 /u_1) c^2 - K_1 u_1)/g$. When $c = 0$, we have $q = - K_1 u_2/(K_1 u_1 + K_2 u_2)$, and when $c\rightarrow \infty$, $q = -K_1 u_2/(K_1 u_2 + K_2 u_1)$. 
There are several special conditions for the value of $q$: (1) When $u_1 = u_2 = u$, 
$q = - K_1/(K_1 + K_2)$, the two systems can be decoupled into two independent fields similar to that in the symmetric condition. (2) When $c^2 = u_1^2$ or $c^{2}=u_{2}^{2}$, $\theta=n\pi$, where $n\in \mathbb{Z}$, and $q$ can be an arbitrary real number, then $\theta_{1}$ and $\theta_{2}$ are constants. (3) When the denominator equals zero, with $q \rightarrow \infty$, there is no $q$ can fit the equation. Therefore, by Eq. \ref{eq-q}, and from $c^2 \ge 0$, we can find $q \in (-\infty, q_c^1)\cup (q_c^2, +\infty)$, with $q_c^{1,2} = \{-\frac{K_{1} u_{2}}{K_{2} u_{1}+K_{1} u_{2}}, -\frac{K_{1} u_{1}}{K_{1} u_{1}+K_{2} u_{2}}\}$. Eq. \ref{eq-thetapp} can be regarded as
the solution of the pendulum in a ring; see a physical interpretation in Fig. S2 of supplemental material S-II \cite{supplematRao}.

It is well-known that the sine-Gordon model supports various solutions \cite{remoissenet2013waves,drazin1989solitons,Ryogo1972Exact,ablowitz1973method}, and one of the possible solutions, termed as a traveling wave  solution, can be written as 
\begin{equation}
    \theta=\pm2\text{am} ( \frac{\sqrt{Ac_{1}-2}\xi}{2\sqrt{A}}\mid-\frac{4}{Ac_{1}-2} ).
    \label{eq-amplitude}
\end{equation}
Here, am means Jacobi amplitude function and $c_{1}$ is a constant. A possible constant in this traveling wave solution has been absorbed into $\xi$. The corresponding density of the system can be obtained using Eq. \ref{eq-conjugatepairs}, giving 
\begin{equation}
    \frac{\delta n_{i}}{n_{i}^{0}}=
    \pm \frac{cqK_{i}}{u_{i}n_{i}^{0}} \sqrt{\frac{Ac_{1}-2}{A}} \operatorname{dn}(
{\sqrt{Ac_{1}-2} \xi \over 2\sqrt{A}} \mid {4 \over 2-Ac_{1}}). 
\end{equation}
From the properties of elliptic functions, $|\text{dn}(x|a)| \le 1$, $a=-4/(Ac_{1}-2)$, thus the magnitude of $\delta n_i$ is determined by $K_{i}/\sqrt{A}$. The estimation of the magnitude of $\delta n_i$ in experiments will be conducted later.   

\begin{figure}[htbp]
\centering
\includegraphics[width=0.45\textwidth]{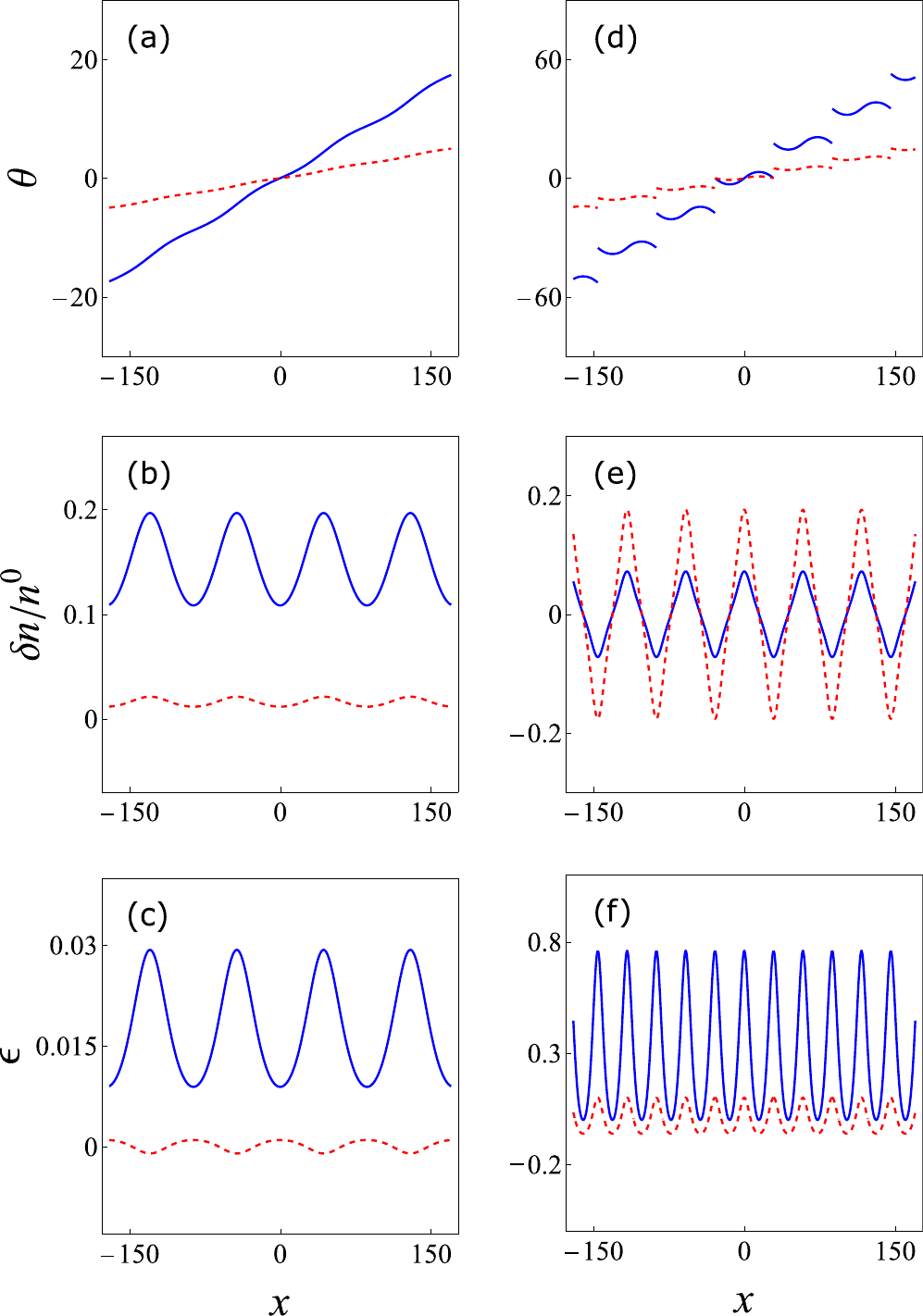}
\caption{The phase, particle density, and energy density in two typical examples (left and right). Parameters used in all figures are $ K_{1} = 1.5$, $K_{2} = 2.5$,
 $u_{2} = 1.3$, $q = 0.4$, $t=0$, $\mu_{1}=0.5$, $\mu_{2}=0.5$. In (a) and (b) $u_{1}=0.8$,  $c_{1}=0.006$, $g=0.001$; and in (d) and (e),  $u_{1}=3.8$, $c_{1}=0.03$, $g=0.1$; and (c) and (f) show the corresponding energy density $\mathcal{T}$ (blue solid line), $\mathcal{V}$ (red dashed line). }
\label{fig-fig2}
\end{figure}

We can obtain the energy density based on this solution. We present some typical traveling wave solutions in Fig. \ref{fig-fig2}, in which the phase can be a continuous function or a discrete function depending on the value of $-4/(Ac_1 -2)$. Since the phase appears in the density of energies by $\partial \theta$ and $\cos(\theta_1 - \theta_2)$, the discreteness of the phase will not introduce extra energy to these traveling wave solutions. These results also show the possible observation of them in experiments due to the finite density of energy in these conjugate pairs. 

(II) {\it Coherent soliton solution}. This generalized sine-Gordon model is most intriguing for the various possible solutions presented, in which the most widely studied one is the following kink and anti-kink solution
\begin{equation}
\theta=4\arctan(\exp(\pm\frac{x+ct+\theta_{0}}{\sqrt{A}})), \quad A>0.
\label{eq-artctan}
\end{equation}
This solution can also be obtained by the integral of motion method \cite{Yomosa1983soliton}, fixing the boundary condition that 
$\theta(+\infty) - \theta(-\infty) = 2n\pi$, $n\in \mathbb{Z}$. 
When $A <0$, a phase shift of $-\pi$ should be introduced to the above solution. This solution can also be obtained from the Jacobi functions using $\text{dn}(x|1) = \sech(x)$, and 
$\text{am}(x|1) = 2\text{arctan}(e^x) - \pi/2$. So Eq. \ref{eq-artctan} can be considered as a special case of Eq. \ref{eq-amplitude} with infinite period; see more details in supplemental material S-III \cite{supplematRao}. Choosing the plus sign in the bracket, the density is given by 
\begin{equation}
{\delta n_{1} \over n_1^0} =-\frac{2c(1+q)K_{1}\sech\left(\frac{x+ct+\theta_{0}}{\sqrt{A}}\right)}{\sqrt{A}u_{1} n_i^0},
\label{eq-sech}
\end{equation}
and $\delta n_{2}$ can be obtained similarly.  This model also supports other types of solutions. For example, Eq. \ref{eq-thetapp} can be found in a lot of conditions. It is a standard solution for the pendulum and Josephson junction \cite{Landau1976Mechanics}, which can be solved exactly, yielding periodic and soliton solutions. It may also support multiple solitons, which will not be discussed in this work \cite{Ryogo1972Exact,ablowitz1973method}. Some more intriguing solutions of this sine-Gordon model are presented in Ref. \cite{He2012non}.

\begin{figure}[htbp]
\centering
\includegraphics[width=0.45\textwidth]{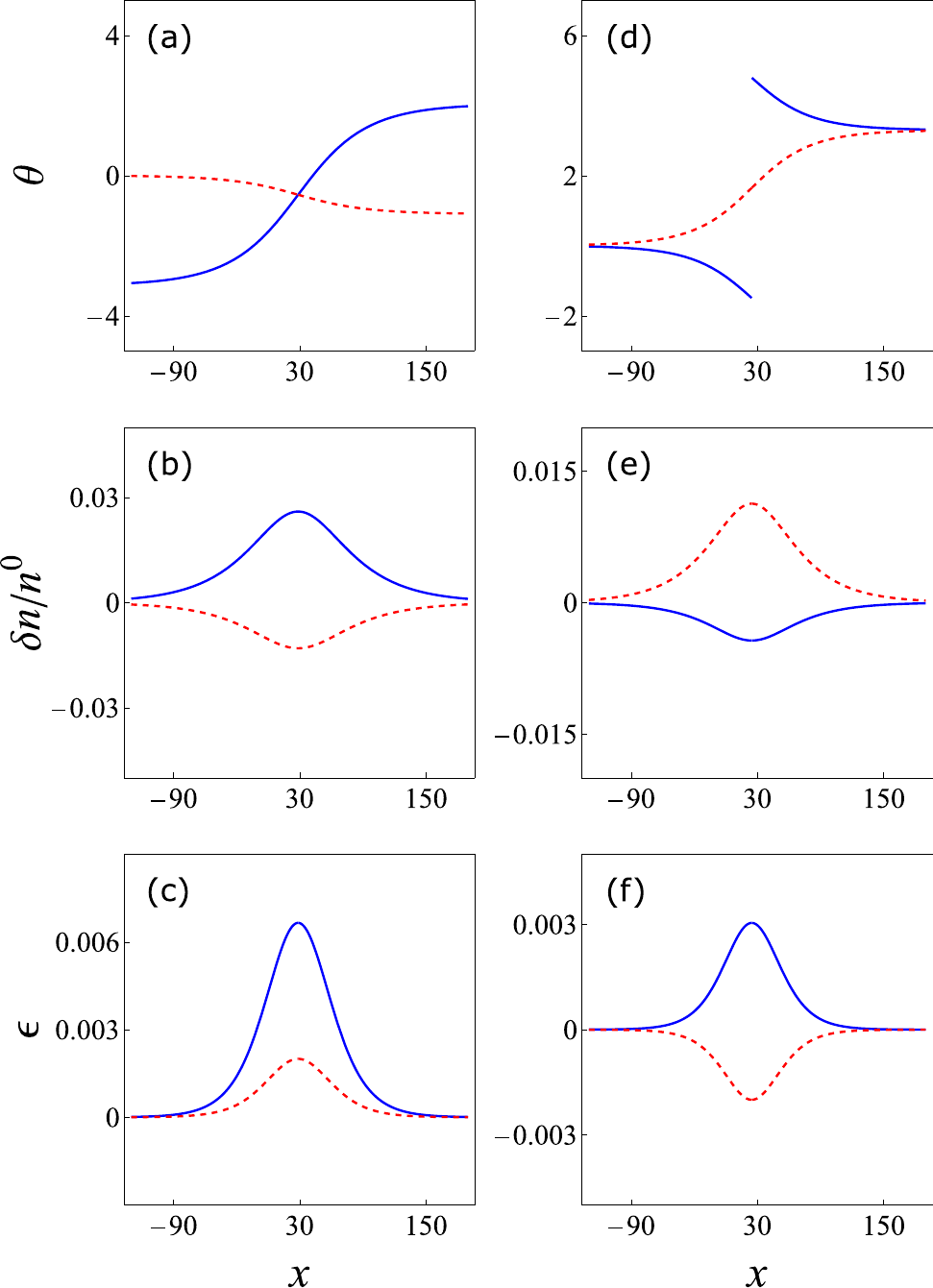}
\caption{The same as that for Fig. \ref{fig-fig2} for two typical examples. Parameters
used in all figures are $ K_{1} = 1.5$, $K_{2} = 2.5$, $u_{1}=2$, $u_{2} = 1.3$, $k=1$, $g = 0.001$, $t=0$. In (a) and (c), $c=2$; and in (d) and (f), $c=0.5$; and (c) and (f) show the corresponding energy density $\mathcal{T}$ (blue solid line), $\mathcal{V}$ (red dashed line).}
\label{fig-fig3}
\end{figure}

(III) {\it Incoherent soliton solution}. Different from  (II), this model may also support incoherent solitons, which means that the two phase fields may have totally different behaviors; thus, $\theta_1$ and $\theta_2$ will depend on $\xi$ in a complicated and different way. To solve this equation, we define $\varphi=\theta_{2}-\theta_{1}$ for convenience. Let \cite{SALAS2010Exact, EKICI2017Exact, Aliyu2021dynamics}
\begin{equation}
    \theta_{2}=\theta(\xi), \quad 
    \varphi = 2 \tan^{-1}(v(\xi)),
 \end{equation}
where $\beta$ and $c$ are constants and $\xi = \beta(x - ct)$. We have the following two equivalent solutions based on Eq. \ref{eq-motion1} and Eq. \ref{eq-motion2} that 
\begin{equation}
    \theta^{\prime\prime}(\xi)=-\frac{2\delta^{2}v(\xi)}{(c^{2}-1)\beta^{2}(v^{2}(\xi)+1)},
\end{equation}
with $v = v(\xi)$ is subjected to 
$(v^2+1) v'' -2 v(v')^2 + \gamma (v^3 + v) =0$, where $\gamma = ((c^2-\alpha^2)\delta^{2}+c^2-1)/((\beta^2 (c^2 - \alpha^2)(c^2-1)))$ is a constant. This solution has a beautiful property: if $v$ is its solution with $\gamma$, then so is $u = 1/v$ with $\gamma \rightarrow -\gamma$, which can be used to construct the singular soliton. Let $v = \tan(\varphi/2)$ then  
\begin{equation}
\varphi'' = -\gamma \sin(\varphi),
\label{eq-varphi}
\end{equation}
which is the sine-Gordon model in Eq.  \ref{eq-thetapp}. We obtain
\begin{eqnarray}
v_1(\xi) && = \pm(-k \mathrm{e}^{-\xi}+\frac{1}{4 k} \mathrm{e}^{\xi}), \quad \gamma=1, \\ 
v_2(\xi) && = \pm \frac{4 k}{ \mathrm{e}^{\xi}-4 k^2\mathrm{e}^{-\xi}} =1/v(\xi), \quad \gamma=-1.
\end{eqnarray}
The corresponding solutions of the $\theta$ field read as 
\begin{equation}
\theta  =-\frac{4 \delta^2}{\left(c^2-1\right) \beta^2} \cot ^{-1}(2 k \exp (-\xi))+C_1 \xi+
C_{2},
\end{equation}
where $C_{1}$, $C_{2}$, $k$ are constants. The constants $C_1$, $C_{2}$ will not enter the total energy $E$. In this work, we only present the results with $C_1 = C_2=0$. With the inverse transformation $t'\rightarrow t, x'\rightarrow x$, the corresponding density for $v_1$ can be obtained from
\begin{eqnarray}
    \delta n_{1} && =\sqrt{\frac{gK_{1}}{u_{1}}}\frac{8ck(\beta^{2}(c^{2}-1)-\delta^{2})Q}{\beta(c^{2}-1)(Q^{2}+4k^{2})}, \\ 
    \delta n_{2} && =-\frac{K_{2}}{u_{2}}\sqrt{\frac{gu_{1}}{K_{1}}}\frac{8c\delta^{2}kQ}{\beta(c^{2}-1)(Q^{2}+4k^{2})},
\end{eqnarray}
where $Q= \exp(\beta \sqrt{\frac{gu_{1}}{K_1}}\left(x-c u_2 t\right)/u_2)$. It should be noted that we have transferred $x$ and $t$ to their original form, thus, $c$ is a dimensionless constant. The other solutions for $v_2(\xi)$ in a similar form can be found in the supplemental material S-IV \cite{supplematRao}. With these expressions, the corresponding phases, densities, and energies are presented in Fig. \ref{fig-fig3}. 

(IV) {\it Experiments observations}. Finally, let us discuss the feasibility of the experimental observation of these solutions. (1) The two coupled chains can be realized using the experiments in 
\cite{Ji2022Floquet, haller2010pinning,  schweigler2017experimental, schweigler2021decay}. It can also be realized using the two-component BEC \cite{rao2023generalized}. (2) The phase difference between the two chains can be measured using the matter-wave interference method \cite{inouye1999phase,Ji2022Floquet}; and the density difference between the two chains can be measured using time-of-flight imaging. (3) The key parameter to distinguish it is the contrast $\mathcal{R}_i = \delta n_{i}/ n_i^0$, which needs to be estimated using experimental parameters \cite{inouye1999phase,Ji2022Floquet,Gritsev2007Linear,Cazalilla2004Bosonizing}. We set $c=\alpha u_{i}$ and $\alpha\sim 1$, then $\delta n_{i}\sim \frac{K_{i}}{\sqrt{A}}\sim\sqrt{\frac{gK_{i}}{u_{i}}}\sim \sqrt{\frac{2Jn_{i}^{0}}{U_{i}}}$ 
is valid in both of these solutions, and the healing length of the soliton is estimated to be $\xi\sim \sqrt{A}\sim\frac{\hbar}{\sqrt{2}}\frac{1}{\sqrt{Jm}} $. The interaction between atoms confined in a 1D waveguide with transverse harmonic confinement
is well described by a zero-range interaction potential $v(x-x^{\prime})=U\delta(x-x^{\prime})$ with \cite{Gritsev2007Linear,Cazalilla2004Bosonizing}
\begin{equation}
U=\frac{2 \hbar^2}{M \ell_{\perp}} \frac{a_s / \ell_{\perp}}{1-C\left(a_s / \ell_{\perp}\right)},
\end{equation}
where $a_s$ is the three-dimensional scattering length, $C=1.0325 \ldots$, and $\ell_{\perp}=\sqrt{\hbar / M \omega_{\perp}}$ is the transverse oscillator length. 
To evaluate $\delta n$, we choose $J=h\times 23$ 
Hz, $\omega_{\perp}=2\pi\times 1.4$ kHz, $n_{0}=50$ $\mu\text{m}^{-1}$, $a_{s} \sim 1-10$ nm, 
$l_{\perp}\sim 72$ nm \cite{Ji2022Floquet}. With these parameters, we have $\delta n_{i}\sim 1$ $\mu\text{m}^{-1}$ and $\xi\sim 1-10$ $\mu \text{m}$, which satisfies the assumption $\delta n_{i}\ll n_{0}$. Here, both the densities and phases can be directly measured in experiments; however when $\delta n_{i}$ is too small to be distinguished, we can probe the phase difference between the two condensates, which has been realized in  experiments  \cite{haller2010pinning,schweigler2017experimental,schweigler2021decay}, and then extract the corresponding density fluctuation using the conjugate relation in Eq. \ref{eq-conjugatepairs}.

To conclude, in this work, we demonstrate that the coupled 1D BECs, in which the excitations are described by two coupled sine-Gordon equations, can provide a fertile ground for exploring various soliton and traveling wave physics. We utilize the fact that the phase and density fields are conjugate parts, and by using the density-phase representation and integrating out the density field, we can obtain effective equations for the two phase fields. We unveil two major solutions, while much more intriguing solutions, including multiple solitons (train) and kink anti-kink physics, can also be realized in this system. The framework used in this work exemplifies the unique feature of the 1D system, thus, it is hoped that these results can find applications in the future for exploring soliton physics in ultracold atoms, which may also be realized using ultracold Fermion atoms. 

In the three-dimensional ultracold atom systems, the fluctuation effect can be taken into account using the Lee-Huang-Yang (LHY) correction, which has recently been used to explain the
stability of self-bound droplet \cite{Chomaz2016quantum,Skov2021Observation,Jorgensen2018Dilute,Lavoine2021Beyond,Petrov2015Quantum, cabrera2018quantum,luo2021new, Abbas2024Selfbound}. This correction can be obtained by the quasi-particle excitation, assuming that the phase difference between the two chains is small enough, hence the sine term in sine-Gordon equation can be approximated using $\sin(\theta_1 - \theta_2) \simeq \theta_1 - \theta_2$; see Ref. \cite{Whitlock2003Relative}. Furthermore, this approximation may lead to infrared divergence of non-condensate density, yielding failure of mean-field theory \cite{Henkel2017Crossover}. In the exact 1D systems without quasi-particle excitation, this approximation may not be applicable anymore, and the method employed in this work and the related sine-Gordon model is much more suitable to describe these physics. Recently, this effective model has been realized in experiments \cite{haller2010pinning, schweigler2017experimental, schweigler2021decay}, thus, we expect the solutions and the associated models in this work, which are beyond the mean-field paradigm in 1D systems, to be realized with cold atoms in the near future.  

\textit{Acknowledgments}: This work is supported by the Strategic Priority Research Program of the Chinese Academy of Sciences (Grant No. XDB0500000)  and the Innovation Program for Quantum Science and Technology (2021ZD0301200, 2021ZD0301500). 

\bibliography{ref.bib}

\end{document}


\title{Supplemental Material to  “Soliton and traveling wave solutions in coupled one-dimensional condensates”}
\author{Zeyu Rao}
\affiliation{Key Laboratory of Quantum Information, University of Science and Technology of China, Hefei 230026, China}
\author{Xiaoshui Lin}
\affiliation{Key Laboratory of Quantum Information, University of Science and Technology of China, Hefei 230026, China}
\author{Jingsong He}
\affiliation{Institute for Advanced Study, Shenzhen University, Shenzhen, 518000, China}
\affiliation{College of Physics and Optoelectronic Engineering, Shenzhen University, 518060, Shenzhen, China}
\author{Guangcan Guo}
\affiliation{Key Laboratory of Quantum Information, University of Science and Technology of China, Hefei 230026, China}
\affiliation{Hefei National Laboratory, University of Science and Technology of China, Hefei 230088, China}
\affiliation{Synergetic Innovation Center of Quantum Information and Quantum Physics, University of Science and Technology of China, Hefei 230026, China}
\affiliation{Anhui Province Key Laboratory of Quantum Network,
University of Science and Technology of China, Hefei 230026, China}
\author{Ming Gong}
\email{gongm@ustc.edu.cn}
\affiliation{Key Laboratory of Quantum Information, University of Science and Technology of China, Hefei 230026, China}
\affiliation{Hefei National Laboratory, University of Science and Technology of China, Hefei 230088, China}
\affiliation{Synergetic Innovation Center of Quantum Information and Quantum Physics, University of Science and Technology of China, Hefei 230026, China}
\affiliation{Anhui Province Key Laboratory of Quantum Network,
University of Science and Technology of China, Hefei 230026, China}

\date{\today }
\maketitle

\renewcommand{\thefigure}{S\arabic{figure}}
\renewcommand{\theequation}{S\arabic{equation}}

\section{Solitons for Gross-Pitaevskii Equation}
\label{sec-solitongp}

The dynamics of one-dimensional (1D) Bose-Einstein condensate (BEC) can be described by the following Gross-Pitaevskii (GP) equation \cite{Pitaevskii2016BEC,Tsuzuki1971Nonlinear,Morsch2006Dynamics,Dalfovo1999TheoryBEC,Konotop2004Landau}
\begin{equation}
i\hbar{\partial \psi (x, t)\over \partial t} = (-{\frac{\hbar^{2}}{2m}{\partial^2 \over \partial x^2} + g|\psi(x, t)|^2}) \psi(x, t),
\end{equation}
where $m$ is the atom mass, $\hbar$ is the reduced Planck constant divided by $2\pi$ and $g$ is the many-body interaction strength. When $g<0$, it admits a self-focusing bright soliton and the above equation can be written as 
\begin{equation}
    i\hbar{\partial \psi (x, t)\over \partial t}+{\frac{\hbar^{2}}{2m}{\partial^2\psi(x,t) \over \partial x^2}}+|g| |\psi(x, t)|^2\psi(x, t)=0.
\end{equation}
With the transformation ($l$ is the typical length scale in this system)
\begin{equation}
    \tilde{t}=\frac{\hbar t}{2m l^2}, \quad \tilde{x}=\frac{x}{l},\quad \tilde{\psi}(\tilde{x},\tilde{t})=\mathcal{A}\psi(\tilde{x},\tilde{t}),
\end{equation}
where $\mathcal{A}$ is a constant making the coefficient of the nonlinear term to be unity. In this way, we have the following dimensionless 1D GP equation
\begin{equation}
    i\frac{\partial\tilde{\psi}(\tilde{x},\tilde{t})}{\partial \tilde{t}}+\frac{\partial^{2}\tilde{\psi}(\tilde{x},\tilde{t})}{\partial \tilde{x}^{2}}+| \tilde{\psi}(\tilde{x},\tilde{t}) |^{2} \tilde{\psi}(\tilde{x},\tilde{t})=0.
\end{equation}
The solitary solution to the above equation is 
\begin{equation}
    \tilde{\psi}(\tilde{x},\tilde{t})=a\exp(i(\frac{c}{2}(\tilde{x}-c\tilde{t})+\frac{1}{2}(a^{2}+\frac{1}{2}c^{2})\tilde{t}))\sech(\frac{a}{\sqrt{2}}(\tilde{x}-c\tilde{t})),
\end{equation}
where $a$ is an arbitrary constant and $c$ is the velocity. The details of solitons for the GP equation can be found in Ref \cite{drazin1989solitons}. If we restrict our consideration to static solitonic solutions, as previously addressed in the main text, we have $c=0$, giving 
\begin{equation}
    \tilde{\psi}(\tilde{x},\tilde{t})=a\exp(\frac{i}{2}a^{2}\tilde{t})\sech(\frac{a}{\sqrt{2}}\tilde{x}).
\end{equation}

When $g > 0$, we should have the following assumption 
\begin{equation}
    \tilde{\psi}(\tilde{x},\tilde{t})= \tilde{\phi}(\tilde{x})e^{-i\tilde{\mu} \tilde{t}}.
\end{equation}
Then the GP equation becomes (we have done the same as the above to obtain the dimensionless equation)
\begin{equation}
    \tilde{\mu} \tilde{\phi}(\tilde{x})=(-{{\partial^2 \over \partial \tilde{x}^2} + |\tilde{\phi}(\tilde{x})|^2}) \tilde{\phi}(\tilde{x}).
\end{equation}
We can obtain the dark soliton solution
\begin{equation}
    \tilde{\phi}(\tilde{x})=\tilde{A} \tanh( \frac{\tilde{A}}{\sqrt{2}} \tilde{x}),\quad \tilde{\mu}=\tilde{A}^{2}.
\end{equation}
To the above results, the following identities may be required 
\begin{equation}
\tanh^2(x) + \sech^2(x) = 1, \quad 
\tanh''(x) = -2\sech^2(x) \tanh(x), \quad 
\sech''(x) = \sech(x) - 2\sech^3(x).
\end{equation}
The schematic diagrams of the dark soliton $\vert \psi(x)\vert^{2}\sim (\sech(x))^{2}$ and the bright soliton $\vert \psi(x)\vert^{2}\sim (\tanh(x))^{2}$; $ \psi(x) \sim \sech(x)$ and kink soliton $ \psi(x) \sim \tanh(x)$ are shown in Fig. 3. One may find more details about solitons and further discussion about the rich non-linear phenomena of BEC in Ref \cite{KIVSHAR1998Dark,Frantzeskakis2010Dark,Carretero2008Nonlinear}.

\begin{figure}[htbp]
\centering
\includegraphics[width=0.8\textwidth]{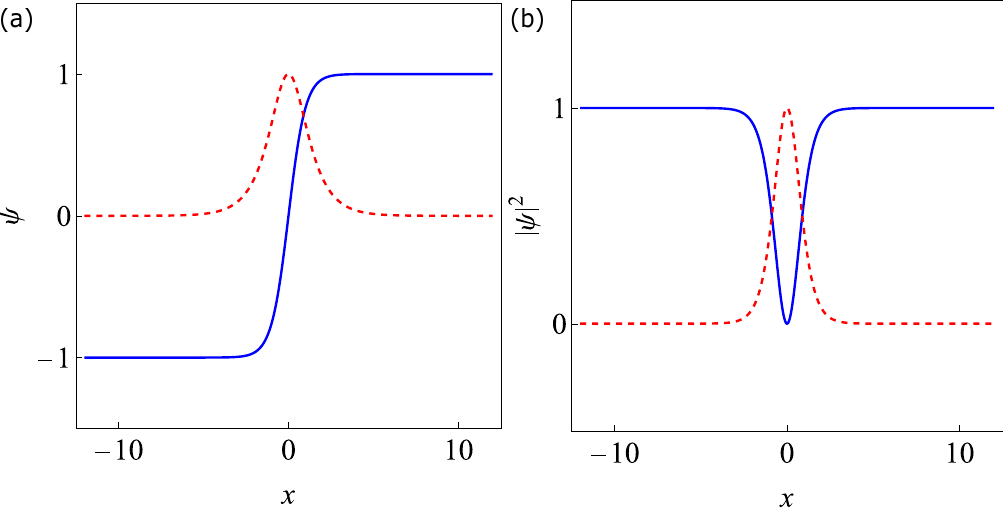}
\label{schematic-diagram}
\caption{(a) Kink soliton $\psi(x) \sim\tanh(x)$ (blue solid line), and bright soliton $\psi(x) \sim\sech(x)$ (red dashed line). (b) For density $\vert \psi(x) \vert^{2}$, dark soliton gives $\vert \psi(x) \vert^{2}=(\tanh(x))^{2}$ (blue solid line) and bright soliton gives $\vert \psi(x) \vert^{2}=(\sech(x))^{2}$ (red dashed line). }
\end{figure}

\section{Lagrangian and Hamiltonian of the two coupled condensates}
\label{sec-Lagrangian}
\begin{figure}[htbp]
\centering
\includegraphics[width=0.8\textwidth]{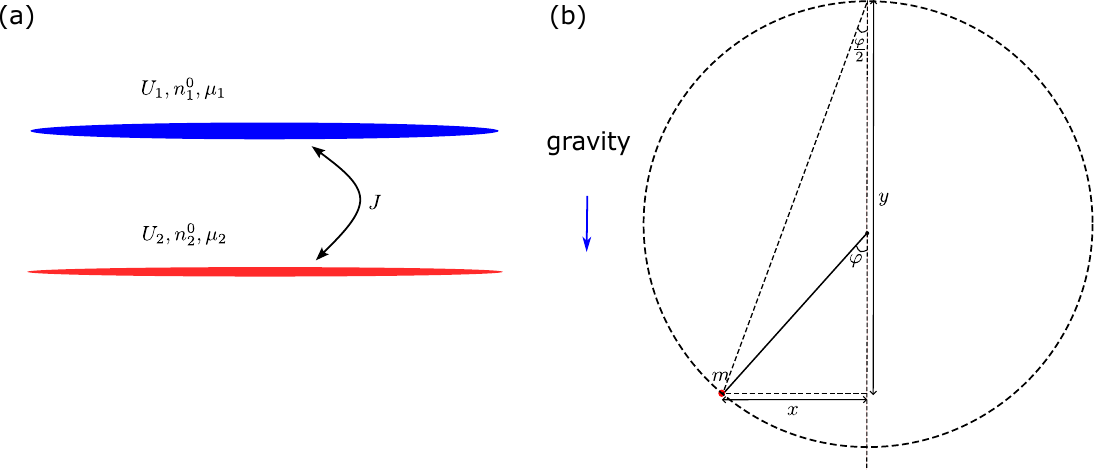}
\label{fig-appendix1}
\caption{(a) Coupled one-dimensional condensates for phase soliton and density soliton. The parameters of $U_{i}$, $ n_{i}^{0}$, and $\mu_{i}$, for $i=1$, $2$, and the inter-chain coupling $J$ are used in the main text. (b) The equivalent pendulum dynamics in a gravity field for the periodic 
oscillations. }
\end{figure}

The density of the Lagrangian and Hamiltonian for the coupled condensates can be written as 
\begin{equation}
\mathcal{L} = \sum_{i=1,2} {i \hbar \over 2} (\psi_i^\dagger \partial_t \psi_i - \psi_i \partial_t \psi_i^\dagger) - \mathcal{H},
\label{eq-effectiveL}
\end{equation}
and 
\begin{equation} 
\mathcal{H} = \sum_{i=1,2} \psi_i^\dagger ({p^2 \over 2m} -\mu_i)\psi_i + {U_i \over 2} (\psi_i^\dagger \psi_i)^2 + J\psi_1^\dagger \psi_2 + \text{h.c.},
\end{equation}
where $\psi_i$ is the bosonic field of the $i$-th chain/condensate with $m$ being the particle mass, $\mu_i$ is the chemical potential, $U_i$ is the many-body interaction strength, and $J$ is the hopping strength between the two condensates and $\hbar$ is the reduced Planck constant. We can represent the bosonic fields using the phase-density representation \cite{Cazalilla2011one,Haldane1981effective,Ruggiero2021LargeScale, giamarchi2003quantum, rao2023generalized}
\begin{equation} 
\psi_i = e^{i\theta_i}\sqrt{n_{i}^{0}+\delta n_{i}} , \quad \psi_i^\dagger =  \sqrt{n_{i}^{0}+\delta n_{i}}e^{- i\theta_i},\quad \psi_{i}^{\dagger}\psi_{i}= n_{i}^{0}+\delta n_{i}. 
\end{equation}
Here, $n_{i}^{0}$ is the average density of the $i$-th condensate 
and $\delta n_i$ is its fluctuation, satisfying $\delta n_{i}\ll n_{i}^{0}$. Hereafter, for convenience, we have dropped the 
arguments in these operators, that is, $A = A(x, t)$. 
We have the following commutation relation using these new variables 
\begin{equation}
    [\delta n_i(x,t), \theta_j(x',t)] = 
i\delta_{ij}\delta(x-x'),\quad [\delta n_{i}(x,t), e^{i\theta_{j}(x',t)}]=-\delta_{ij}\delta(x-x')e^{i\theta_{j}(x',t)}.
\end{equation}

The kinetic energy of this model is related to the following term
\begin{equation} 
\nabla \psi_i = e^{i\theta_i} (\nabla \sqrt{n_{i}^{0}+\delta n_{i}} + i (\nabla \theta_i) \sqrt{n_{i}^{0}+\delta n_{i}}).
\end{equation}
Using the fact that the first term is real and the second term is a pure imaginary number, we have 
\begin{equation}
|\nabla \psi_i|^2 = |\nabla \sqrt{n_{i}^{0}+\delta n_{i}}|^2 + (n_{i}^{0}+\delta n_{i}) |\nabla \theta_i|^{2}.
\end{equation}
The time derivative term for the Legendre transformation in the Lagrangian can be understood similarly, and we have 
\begin{equation} 
\psi_i^\dagger \partial_t \psi_i = \sqrt{n_{i}^{0}+\delta n_{i}} \partial_t \sqrt{n_{i}^{0}+\delta n_{i}} + i (n_{i}^{0}+\delta n_{i}) \partial_t \theta_i,
\end{equation}
in which the first term is a pure real number and the second term is imaginary. Since we have $i$ in front of the Lagrange equation, only the Hermite term survives. Furthermore, from the commutation relation, we find equivalence $x \sim - \theta$ and $p \sim n_i$, so from the standard Legendre transformation from the Lagrangian to the Hamiltonian, it should be $p\dot{x}$, yielding the same result. The Hamiltonian can be calculated as follows
\begin{eqnarray}
    && \psi_{i}^{\dagger}\frac{p^{2}}{2m}\psi_{i}=\frac{\hbar^{2}}{2m}\left[ (n_{i}^{0}+\delta n_{i})(\partial_{x}\theta_{i})^{2} +\frac{(\partial_{x}\delta n_{i})^{2}}{4(n_{i}^{0}+\delta n_{i})}\right], \\ 
    && -\mu_{i}\psi_{i}^{\dagger}\psi_{i}=-\mu_{i}(n_{i}^{0}+\delta n_{i}),\\
    && \frac{U_{i}}{2}(\psi_{i}^{\dagger}\psi_{i})^{2}=\frac{U_{i}}{2}(n_{i}^{0}+\delta n_{i})^{2}, \\ 
    && J\psi_{1}^{\dagger}\psi_{2}+J\psi_{2}^{\dagger}\psi_{1}=2J\sqrt{n_{1}^{0}n_{2}^{0}}( 1+\frac{\delta n_{2}}{2n_{2}^{0}}+\frac{\delta n_{1}}{2 n_{1}^{0}}+\frac{\delta n_{1}\delta n_{2}}{4n_{1}^{0}n_{2}^{0}} )\cos(\theta_{1}-\theta_{2}).
\end{eqnarray} 
Due to $\delta n_{i}\ll n_{i}^{0} $,  the Hamiltonian is given as
\begin{equation}
    \mathcal{H}= \varepsilon_g + 2J\sqrt{n_{1}^{0}n_{2}^{0}}\cos(\theta_{1}-\theta_{2})+\sum_{i=1,2}\frac{1}{2}U_{i}\delta n_{i}^{2}+\frac{\hbar^{2}(\partial_{x}\delta n_{i})^{2}}{8mn_{i}^{0}}+\frac{\hbar^{2}n_{i}^{0}(\partial_{x}\theta_{i})^{2}}{2m}-(\mu_{i}-n_{i}^{0}U_{i})\delta n_{i},
\end{equation}
where 
\begin{equation}
\varepsilon_g = {1\over 2} U_1 (n_1^0)^2 + {1\over 2} U_2  (n_2^0)^2 -\mu_1 n_1^0 - \mu_2 n_2^0.
\end{equation}
To obtain the minimal of $\epsilon_{g}$ we have
\begin{equation}
    \frac{\partial \epsilon_{i}}{\partial n_{i}^{0}}=0.
\end{equation}
The above equation will give 
\begin{equation}
    \mu_{i}=U_{i}n_{i}^{0}.
\end{equation}
In this model, the integration of $\mu_{i}\delta n_{i}$ will become zero, $\int dx \delta n_i = 0$, thus it has no contribution. The value of $n_i^0$ is given by the minimal of $\varepsilon_g$, assuming that $n_1^0 + n_2^0 = n^0$. We find 
\begin{equation}
n_1^0 = {\mu_1 - \mu_2 \over U_1 + U_2} + {U_2 \over U_1 + U_2} n^0, \quad n_2^0 = {\mu_2 - \mu_1 \over U_1 + U_2} + {U_1 \over U_1 + U_2} n^0.
\end{equation}
The excitation part can be split into two parts
\begin{eqnarray}
    \mathcal{T} && = \sum_{i=1}^2  \frac{(\partial_{x}\delta n_{i})^{2}}{8 mn_{i}^{0}}+\frac{n_{i}^{0} (\partial_{x}\theta_{i})^{2}}{2 m}+\frac{1}{2}U_{i}\delta n_{i}^{2}, \label{eq-T} \\ 
    \mathcal{V} && =2J\sqrt{n_{1}^{0}n_{2}^{0}}\cos{(\theta_{1}-\theta_{2})}. \label{eq-V} 
\end{eqnarray}
And the total energy density 
\begin{equation}
\epsilon = \mathcal{T}+\mathcal{V}.
\end{equation}
Here $\mathcal{T}$ is the density of the kinetic energy, and $\mathcal{V}$ is the density of the interaction energy. These energy densities are calculated in Fig. 1 and Fig. 2 in the main text. 

In the above analysis, we have ignored some constant terms in $ \mathcal{H}$. In this way, the Lagrangian density is given as
\begin{equation}
    \mathcal{L}=-\hbar (n_{1}^{0}+\delta n_{1})\partial_{t}\theta_{1}-\hbar (n_{2}^{0}+\delta n_{2})\partial_{t}\theta_{2}-\mathcal{H}.
\end{equation}
Using the canonical relation, we can obtain the relation between $\partial_{t}\theta_{i}$ and $\delta n_{i}$
\begin{eqnarray}
i\hbar\partial_{t}\theta_{i}(x,t) && = 
 [\theta_{i}(x,t),H(x',t)] \nonumber \\ 
 && = [\theta_{i}(x,t),\int \mathcal{H}(x',t)dx'] \nonumber \\
 && = \int[\theta_{i}(x,t),\frac{1}{2}U_{i}(\delta n_{i}(x',t))^{2}]dx' \nonumber \\
 && = \int [-iU_{i}\delta n_{i}(x',t)\delta(x-x')]dx' \nonumber \\
 && = -i U_{i}\delta n_{i}(x,t).
 \label{appendix-conjugate1}
\end{eqnarray}
Thus we have
\begin{equation}
    \delta n_{i}=-\frac{\hbar\partial_{t}\theta_{i}}{U_{i}}.
    \label{appendix-conjugate2}
\end{equation}
We need to mention that we have ignored $\frac{\hbar^{2}(\partial_{x}\delta n_{i})^{2}}{8mn_{i}^{0}}$ in $\mathcal{H}$ when $\delta n_{i}\ll n_{i}^{0}$, in which condition the derivative of $\delta n_{i}$ is much smaller. Physically, the excitation of phase can be gapless; while the excitation of density is fully gapped in the presence of $U_i \delta n_i^2$ term. In this way, the density of the Lagrangian can be written as 
\begin{equation}
    \mathcal{L}=\frac{\hbar^{2}(\partial_{t}\theta_{1})^{2}}{2U_{1}}+\frac{\hbar^{2}(\partial_{t}\theta_{2})^{2}}{2U_{2}}-\frac{\hbar^{2}n_{1}^{0}(\partial_{x}\theta_{1})^{2}}{2m}-\frac{\hbar^{2}n_{2}^{0}(\partial_{x}\theta_{2})^{2}}{2m}-2J\sqrt{n_{1}^{0}n_{2}^{0}}\cos(\theta_{1}-\theta_{2}),
\end{equation}
which is the central equation used in this work. 

The Euler-Lagrange equation of motion for the above Lagrangian can be written as 
\begin{eqnarray}
    \frac{\hbar^{2}\partial_{t}^{2}\theta_{1}}{U_{1}}-\frac{\hbar^{2}n_{1}^{0}\partial_{x}^{2}\theta_{1}}{m} && = + 2J\sqrt{n_{1}^{0}n_{2}^{0}}\sin(\theta_{1}-\theta_{2}) \nonumber \\
    \frac{\hbar^{2}\partial_{t}^{2}\theta_{2}}{U_{2}}-\frac{\hbar^{2}n_{2}^{0}\partial_{x}^{2}\theta_{2}}{m} && =-2J\sqrt{n_{1}^{0}n_{2}^{0}}\sin(\theta_{1}-\theta_{2}). 
\end{eqnarray}
The Luttinger parameter and velocity can be defined as
\begin{equation}
    K_{i}=\hbar\sqrt{\frac{n_{i}^{0}}{mU_{i}}},\quad u_{i}=\sqrt{\frac{n_{i}^{0}U_{i}}{m}},\quad g={2J\over \hbar}\sqrt{n_{1}^{0}n_{2}^{0}}.
\end{equation}
Then it leads to
\begin{eqnarray}
    \frac{K_{1}}{u_{1}}\partial_{t}^{2}\theta_{1}-K_{1}u_{1}\partial_{x}^{2}\theta_{1} && = g\sin(\theta_{1}-\theta_{2}), \nonumber \\
    \frac{K_{2}}{u_{2}}\partial_{t}^{2}\theta_{2}-K_{2}u_{2}\partial_{x}^{2}\theta_{2} && = -g\sin(\theta_{1}-\theta_{2}).
\end{eqnarray}
The above equations are used in the main text.

\section{Solutions of the sine-Gordon equation}
\label{sec-sg}

We consider the solution to the following sine-Gordon equation
\begin{equation}
    \partial_t^2 \theta - v^2 \partial_x^2 \theta = g \sin(\theta).
\end{equation}
We consider the solution that $\theta(x, t) = \theta(x - ct)=\theta(\xi)$, where $c$ is the sound velocity. Then we should have 
\begin{equation} 
(c^2 - v^2) \theta '' = g \sin(\theta).
\label{eq-apdix_SG}
\end{equation}
Let us assume that $g/(v^2 - c^2) > 0$; otherwise, we can change $\theta$ to $\theta + \pi$, leading to the same conclusion. Thus $c$ can take any values, including $c\rightarrow \infty$. This second-order differential equation is similar to Newton's equation discussed in textbooks, and we can solve this problem using the integral of motion method, assuming 
\begin{equation} 
E = {\dot{\theta}^2 \over 2} + {g \over v^2 - c^2} ( 1-\cos(\theta)).
\label{eq-pendulum_energy}
\end{equation}
This is nothing but just the solution of the harmonic oscillator constrained in a circle (see Fig. \ref{fig-appendix1}). For this reason, this model can support various solutions, two of which are presented below. 

1. {\bf traveling wave  solution}. It is related to the elliptic functions. According to Eq. \ref{eq-pendulum_energy}, we can map Eq. \ref{eq-apdix_SG} into the pendulum. With the conservation of energy, we should have
\begin{equation}
    {\dot{\theta}^2 \over 2} + {g \over v^2 - c^2} ( 1-\cos(\theta))={g \over v^2 - c^2} ( 1-\cos(\theta_{0})),
\end{equation}
thus 
\begin{equation}
    \frac{d\theta}{d\xi}=\sqrt{\frac{2g}{v^{2}-c^{2}}(\cos(\theta)-\cos(\theta_{0}))}.
\end{equation}
We can solve this solution using the integral of motion method 
\begin{equation}
   \int \frac{d\theta}{\sqrt{\frac{2g}{v^{2}-c^{2}}(\cos(\theta)-\cos(\theta_{0}))}}=\int d\xi,
\end{equation}
yielding 
\begin{equation}
    \sqrt{\frac{2(v^{2}-c^{2})}{g(1-\cos(\theta_{0}))}}\text{F}[\frac{\theta}{2},\csc^{2}(\frac{\theta_{0}}{2})]=\xi-\xi_{0}.
\end{equation}
The Jacobi amplitude function is related to the elliptic integral of the first kind $F(u,k)$ by
\begin{equation}
    \text{F}(\text{am}(u,k),k)=u,
\end{equation}
leading to
\begin{equation}
    \theta=2\text{am}(\sqrt{\frac{g(1-\cos(\theta_{0}))}{2(v^{2}-c^{2})}}(\xi-\xi_{0})\mid \csc^{2}(\frac{\theta_{0}}{2})).
\end{equation}

By redefining the parameters in this solution, a general form can be obtained 
\begin{equation}
    \theta=\pm2\text{am}  (\frac{\sqrt{(c^{2}-v^{2})c_{1}-2g}(x-ct+c_{2})}{2\sqrt{c^{2}-v^{2}}}\mid \frac{4g}{2g+(-c^{2}+v^{2})c_{1}} ),
    \label{eq-amplitude2}
\end{equation}
which has been used in the main text. The Jacobi amplitude function $\text{am}(x\mid m)$ exhibits two different behaviors depending on the value of $m$. When $m > 1$, it is a periodic function of $x$ with finite amplitude, up to some shift of $2\pi n$, $n \in \mathbb{Z}$. In this case, $\cos(\theta)$ 
is still a smooth and continuous function. However when $m < 1$, it can be written as 
\begin{equation}
\text{am}(x\mid m) = kx + \text{periodic oscillation term},
\end{equation}
with $k$ depends on $m$. When $m = 0$, we have am$(x|0) = x$;
and when $m =1$, we have am$(x|1) = \pi/2 \tanh(x)$, which is the kink solution as presented above. 

\begin{figure}[htbp]
\centering
\includegraphics[width=0.5\textwidth]{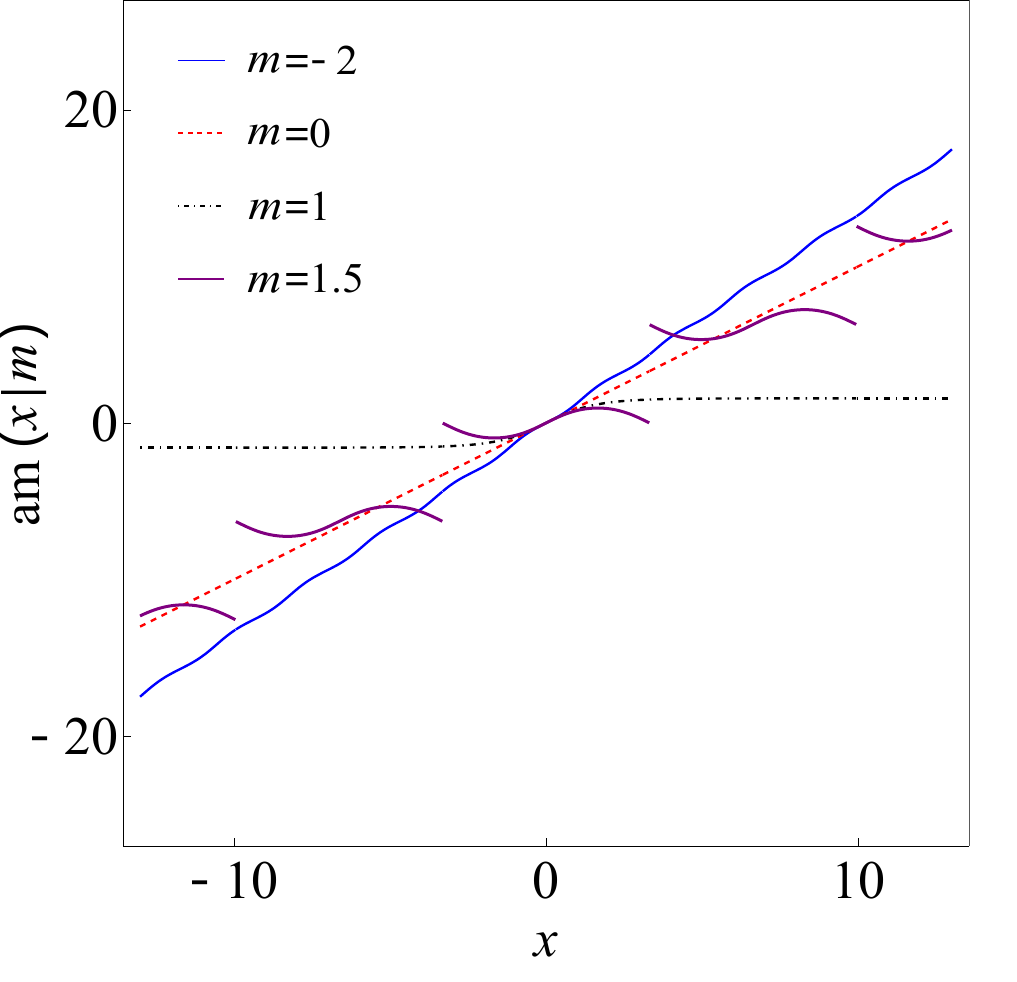}
\label{fig-jacobiam}
\caption{Jacobi amplitude function $\text{am}(x\mid m)$ for various $m=-2$ (blue solid line), $m=0$ (red dashed line), $m=1$ (black dot dashed line), $m=1.5$ (purple solid line).}
\end{figure}

2. {\bf Soliton solution}. When $m = \frac{4g}{2g+(-c^{2}+v^{2})c_{1}}=1$, the above periodic solution reduces to a soliton solution
\begin{equation}
    \theta=\pm  4\arctan(\exp(\frac{\sqrt{(c^{2}-v^{2})c_{1}-2g}(x-ct+c_{2})}{2\sqrt{c^{2}-v^{2}}})) \mp\pi.
    \label{eq-soliton1}
\end{equation}
From the previous discussion, the soliton solution can be regarded as a special case of the elliptic function solution with an infinite period. This soliton solution can also be directly obtained by solving Eq. \ref{eq-apdix_SG}. Multiple $\theta^{\prime}$ on both sides this equation, we obtain 
\begin{equation}
    2\theta^{\prime}\theta^{\prime\prime}=2\frac{g}{c^{2}-v^{2}}\sin(\theta)\theta^{\prime}.
\end{equation}
After integration, it leads to
\begin{equation}
    (\theta^{\prime})^{2}=\frac{4g}{c^{2}-v^{2}}\sin^{2}(\frac{\theta}{2}),
\end{equation}
with the boundary condition\cite{Yomosa1983soliton}
\begin{equation}
    \cos(\theta)=1,\quad \text{when} \quad \xi=\pm\infty. 
\end{equation}
This boundary condition is nothing, but just put the ball on the north pole of the circle $\theta(-\infty) = 0$, which will take infinite time for the ball to come back to its initial condition $\theta(\infty) = 2\pi$. It gives 
\begin{equation}
    \int \frac{d\theta}{2\sin(\frac{\theta}{2})}=\pm\sqrt{\frac{g}{c^{2}-v^{2}}}(\xi-\xi_{0}),
\end{equation}
which can be solved analytically as \cite{Landau1976Mechanics}
\begin{equation}
    \theta=4\arctan(\exp(\pm\sqrt{\frac{g}{c^{2}-v^{2}}}(\xi-\xi_{0}))).
\end{equation}
This solution is identical to Eq.\ref{eq-soliton1} with 
$c_{1}=\frac{6g}{c^{2}-v^{2}}$. We need to mention that with this solution, we can construct the multi-solitons, breather, and degenerate multi-solitons, which are in principle to be realizable in this model \cite{Ryogo1972Exact,ablowitz1973method}. 

\section{Solutions of the coupled sine-Gordon  equation}
\label{sec-coupledsg}

The goal of this section is to solve the following coupled sine-Gordon equation
\begin{eqnarray}
    &&\partial_{t}^{2}\theta_{1}- \alpha^2 \partial_{x}^{2}\theta_{1}=\sin{(\theta_{2}-\theta_{1})},\nonumber \\ &&\partial_{t}^{2}\theta_{2}-\partial_{x}^{2}\theta_{2}=-\delta^2 \sin{(\theta_{2}-\theta_{1})}.
    \label{eq-motion_1}
\end{eqnarray}
Let us define 
\begin{equation}
    \varphi=\theta_{2}-\theta_{1}, \quad \theta_{2}=\theta(\xi),\quad \varphi=2\tan^{-1}(v(\xi)),
\end{equation}
where $\xi=\beta(x-ct)$. From the above definition
\begin{equation}
    \sin{\varphi}=\frac{2v}{1+v^{2}},
\end{equation}
thus the equation of $\theta_{2}$ can be written as
\begin{equation}
    \theta''(\xi)=-\frac{2\delta^{2}v}{\beta^{2}(c^{2}-1)(1+v^{2})}.
    \label{eq-theta_v}
\end{equation}
The solution for the relative phase $\varphi$ and the associated $v$ can be written as 
\begin{equation}
    v''(1+v^{2})-2vv'^{2}+\gamma(v^{3}+v)=0,
\end{equation}
with 
\begin{equation}
    \gamma=\frac{(c^{2}-\alpha^{2})\delta^{2}+c^{2}-1}{\beta^{2}(c^{2}-\alpha^{2})(c^{2}-1)}.
\end{equation}
Furthermore, using $v  = \tan(\varphi/2)$, we find that the above equation is reduced to 
\begin{equation}
{1\over 2} \text{sec}({\varphi \over 2})^4 (\varphi'' + \gamma \sin(\varphi)) = 0, \quad \varphi'' + \gamma \sin(\varphi) = 0. 
\label{eq-reducedvarphi}
\end{equation}
This solution is identical to the solution of the sine-Gordon equation. Using the solution of $\varphi$, we can calculate $v$, and then calculate $\theta(\xi)$.  With the many solutions of $\varphi$, some of which have been discussed in the main text, we also expect various solutions for $\theta$. For example, let us choose $v=v_{1}(\xi)$ \cite{SALAS2010Exact}
\begin{equation}
v_1(\xi)= \pm(-k \mathrm{e}^{-\xi}+\frac{1}{4 k} \mathrm{e}^{\xi}),\quad \gamma=1,
\end{equation}
where $k$ is a constant. If we choose the minus solution, we have 
\begin{equation}
    \theta^{\prime}(\xi)= \int \frac{(e^{\xi}-4e^{-\xi}k^{2})\delta^{2}}{2k\beta^{2}(c^{2}-1)(1+\frac{(e^{\xi}-4k^{2}e^{-\xi})^{2}}{16k^{2}})} d\xi=-\frac{8e^{\xi}k\delta^{2}}{(c^{2}-1)(e^{2\xi}+4k^{2})\beta^{2}}+c_{1},
\end{equation}
thus 
\begin{equation}
    \theta(\xi)=\int -\frac{8e^{\xi}k\delta^{2}}{(c^{2}-1)(e^{2\xi}+4k^{2})\beta^{2}} d\xi=-\frac{4\delta^{2}\cot^{-1}(2ke^{-\xi})}{\beta^{2}(c^{2}-1)}+c_{1}\xi+c_{2}.
\end{equation}
The other solution with plus sign in $v_1(\xi)$ can also be calculated in a similar way. We find that when $\varphi$ is the solution of the sine-Gordon equation with $\gamma$, $\varphi + \pi$ is the solution of the sine-Gordon equation with parameter $-\gamma$. This is reflected from the fact that if we define $\tilde{v} =1/v$, then 
\begin{equation}
    v'=-\frac{\tilde{v}'}{\tilde{v}^{2}},\quad v''=-\frac{\tilde{v}''}{\tilde{v}^{2}}+\frac{2\tilde{v}'^{2}}{\tilde{v}^{3}},
\end{equation}
and the original equation of $v$ becomes
\begin{equation}
    (\frac{1}{\tilde{v}^{2}}+1)(-\frac{\tilde{v}''}{\tilde{v}^{2}}+\frac{2\tilde{v}'^{2}}{\tilde{v}^{3}})-2\frac{\tilde{v}'^{2}}{\tilde{v}^{5}}+\gamma(\frac{1}{\tilde{v}^{3}}+\frac{1}{\tilde{v}})=0.
\end{equation}
The following three steps are usjustifyinged to obtain the equations of $\tilde{v}$ as 
\begin{eqnarray} 
(1+\tilde{v}^{2})(-\tilde{v}\tilde{v}''+2\tilde{v}'^{2})-2\tilde{v}'^{2}+\gamma(\tilde{v}^{2}+\tilde{v}^{4})&& =0, \\ 
-\tilde{v}\tilde{v}''(1+\tilde{v}^{2})+2\tilde{v}^{2}\tilde{v}'^{2}+\gamma(\tilde{v}^{2}+\tilde{v}^{4}) && =0, \\  \tilde{v}''(1+\tilde{v}^{2})-2\tilde{v}\tilde{v}'^{2}-\gamma(\tilde{v}^{3}+\tilde{v})&& =0.
\end{eqnarray} 
So with the solution of $v_{1}(\xi)$, we can get $v_{2}(\xi)=\tilde{v}_{1}(\xi)=1/v_{1}(\xi)$ as following 
\begin{equation}
v_2(\xi)= \pm \frac{4 k}{ \mathrm{e}^{\xi}-4 k^2\mathrm{e}^{-\xi}},\quad\gamma=-1.
\end{equation}

In the following, let us choose the minus solution of $v_1$ and $v_2$, and we have 
\begin{eqnarray}
\theta && =-\frac{4 \delta^2}{\left(c^2-1\right) \beta^2} \cot ^{-1}(2 k \exp (-\xi))+C_1 \xi+C_{2}, \quad \text{for}
\quad v_{1}(\xi) \quad \text{and} \quad v_{2}(\xi); \nonumber \\
\varphi_{1} && = -2 \tan ^{-1}(\frac{1}{4 k}(\exp (\xi)-4 k^2 \exp (-\xi))),\quad \text{for}\quad v_{1}(\xi);\nonumber\\
\varphi_{2} && = -2 \tan^{-1}(\frac{4k}{\exp(\xi)-4k^{2}\exp(-\xi)}),\quad \text{for}\quad v_{2}(\xi),
\label{eq-theta1}
\end{eqnarray}
where $C_{1},C_{2},k$ are arbitrary constants. With Eq. \ref{appendix-conjugate2}, the density component can be obtained as 
\begin{equation}
\delta n_{1}= \frac{8 cgk(\beta^2(c^2-1)-\delta^2) e^{\frac{\beta \sqrt{\frac{gu_{1}}{K_1}}(x-c t u_2)}{u_2}}}{\beta(c^2-1) \sqrt{\frac{gu_{1}}{K_1}}(e^{\frac{2 \beta \sqrt{\frac{gu_{1}}{K_1}}(x-c t u_2)}{u_2}}+4 k^2)},\quad \text{for}\quad v_{1}(\xi),
\end{equation}
\begin{equation}
\delta n_{1}=-\frac{8 cgk(\beta^2(c^2-1)+\delta^2) e^{\frac{\beta \sqrt{\frac{gu_{1}}{K_1}}(x-c t u_2)}{u_2}}}{\beta(c^2-1) \sqrt{\frac{gu_{1}}{K_1}}(e^{\frac{2 \beta \sqrt{\frac{gu_{1}}{K_1}}(x-c t u_2)}{u_2}}+4 k^2)},\quad \text{for}\quad v_{2}(\xi),
\end{equation}
and 
\begin{equation}
\delta n_{2}=-\frac{8 c \delta^2 k K_2 \sqrt{\frac{gu_{1}}{K_1}} e^{-\beta(\frac{x \sqrt{\frac{gu_{1}}{K_1}}}{u_2}-c t \sqrt{\frac{gu_{1}}{K_1}})}}{\beta(c^2-1) u_2(4 k^2 e^{-2 \beta(\frac{x \sqrt{\frac{gu_{1}}{K_1}}}{u_2}-c t \sqrt{\frac{gu_{1}}{K_1}})}+1)},\quad \text{for}
\quad v_{1}(\xi) \quad \text{and} \quad v_{2}(\xi).
\end{equation}
For simplicity, let us define (as used in the main text)
\begin{equation}
    Q= \exp(\beta \sqrt{\frac{gu_{1}}{K_1}}(x-c u_2 t)/u_2),
\end{equation}
then the solution of density can be written as 
\begin{eqnarray}
    \delta n_{1} && =\sqrt{\frac{gK_{1}}{u_{1}}}\frac{8ck(\beta^{2}(c^{2}-1)-\delta^{2})Q}{\beta(c^{2}-1)(Q^{2}+4k^{2})}, \\ 
    \delta n_{2} && =-\frac{K_{2}}{u_{2}}\sqrt{\frac{gu_{1}}{K_{1}}}\frac{8c\delta^{2}kQ}{\beta(c^{2}-1)(Q^{2}+4k^{2})},
\end{eqnarray}
for $v_{1}(\xi)$. 

The solutions for $v_{2}(\xi)$ can be obtained similarly. These solutions of phase and density are used in the main text, showing that when there are some spatial and time dependent structures in the phase, one can also find some corresponding structures in the density component, which is the key idea to be presented in this work. We have chosen some experimental parameters to demonstrate their potential observability, showing that 
\begin{equation}
\delta n_i/n_i^0 \sim 0.1 \ll 1,
\end{equation}
justifying the approximation this supplemental material Sec. \ref{sec-Lagrangian}. 

\bibliography{ref.bib}